\documentclass[journal,draftcls,onecolumn,12pt,twoside]{IEEEtranTCOM}

\usepackage{graphics}
\usepackage{graphicx}
\usepackage{epsfig,amssymb}
\usepackage{amsmath,mathtools}
\usepackage{times}
\usepackage{mathrsfs}
\usepackage{array}
\usepackage{amsmath, latexsym, amsfonts, amssymb}
\usepackage[mathscr]{eucal}
\usepackage{array, tabularx}
\usepackage[table]{xcolor}
\usepackage{psfrag}
\usepackage{multirow}\usepackage{booktabs}
\usepackage{epsfig}
\usepackage{cite}
\usepackage{tikz}
\usepackage{scalerel}
\usepackage{url}
\usepackage{xcolor}
\usepackage[english]{babel}
\usepackage[autostyle]{csquotes}
\usepackage{authblk}
\usepackage{subcaption}
\usepackage{color}

\usepackage{algpseudocode}
\usepackage{algorithm}

\newcommand{\paren}[1]{\left(#1\right)}
\newcommand{\sqparen}[1]{\left[#1\right]}
\newcommand{\brparen}[1]{\left\{#1\right\}}

\newcommand{\PR}[1]{\ensuremath{\mathsf{Pr}\left\{#1\right\}}} 
\newcommand{\ie}{\ensuremath{{\text{\em i.e.}}}}

\newtheorem{lemma}{Lemma}

\begin{document}
%
\title{A Discrete-Time Markov Chain Based Comparison of the MAC Layer Performance of C-V2X Mode 4 and IEEE 802.11p}
%
%
%

\author{Geeth~P.~Wijesiri N.B.A,~\IEEEmembership{Member,~IEEE,}
        Jussi Haapola,~\IEEEmembership{Member,~IEEE,}
        and~Tharaka~Samarasinghe,~\IEEEmembership{Senior Member,~IEEE,}
\thanks{G. P. Wijesiri N.B.A is with the Department of Electronic and Telecommunication Engineering, University of Moratuwa, Sri Lanka, the Center for Wireless Communication, University of Oulu, Finland, and the Department of Electrical and Information Engineering, University of Ruhuna, Sri Lanka (e-mail: geeth@eie.ruh.ac.lk)

J. Haapola is with the Center for Wireless Communication, University of Oulu, Finland (e-mail: jussi.haapola@oulu.fi).

T. Samarasinghe is with the Department of Electronic and Telecommunication Engineering, University of Moratuwa, Sri Lanka, and the Department of Electronic and Electrical  Engineering, University of Melbourne,
Australia (e-mail: tharakas@uom.lk).
}\thanks{The material in this paper was presented in part at the IEEE 90th Vehicular Technology Conference, Hawaii, USA, September 2019 \cite{c_v2xmode4cp}.}}

%
%


\markboth{IEEE Transactions on Communications}%
{Submitted paper}
%



\bibliographystyle{IEEEtranTCOM}
\maketitle
\vspace{-2.5cm}

\begin{abstract}
\vspace{-.6cm}
Vehicle-to-vehicle (V2V) communication plays a pivotal role in intelligent transport systems (ITS) with cellular-vehicle to everything (C-V2X) and IEEE 802.11p being the two competing enabling technologies. This paper presents multi-dimensional discrete-time Markov chain (DTMC) based models to study the medium access control (MAC) layer performance of the IEEE 802.11p standard and C-V2X Mode 4.  
These models are coupled with an appropriate DTMC based queuing model, and traffic generators for periodic cooperative awareness messages (CAMs) and event-driven decentralized environmental notification messages (DENMs). Closed-form solutions for the steady-state probabilities of the models are obtained, which are then utilized to derive expressions for several key performance metrics. An application for a highway scenario is presented to provide numerical results and to draw insights on the performance. In particular, a performance comparison between IEEE 802.11p and C-V2X Mode 4 in terms of the average delay, the collision probability, and the channel utilization is presented.
The results show that IEEE 802.11p is superior in terms of average delay, whereas C-V2X  Mode 4 excels in collision resolution. The paper also includes design insights on possible future MAC layer performance enhancements of both standards.
\end{abstract}


\vspace{-0.75cm}
\begin{IEEEkeywords}
\vspace{-0.75cm}
C-V2X Mode 4, discrete-time Markov chain, ETSI ITS-G5, IEEE 802.11p, medium access control, vehicle-to-vehicle communication.
\end{IEEEkeywords}
\vspace{-0.5cm}

%
\IEEEpeerreviewmaketitle

\vspace{-0.35cm}
\section{Introduction}
\vspace{-0.25cm}
%
%
%
%

\IEEEPARstart{V}{ehicular} networks primarily depend on vehicle-to-everything (V2X) communications
in enabling an active safety environment. V2X communication mainly consists of an exchange of small broadcast packets that have critical latency and reliability constraints. 
IEEE 802.11p/ dedicated short-range communication (DSRC) is known to be the first commercial V2X communication technology.
As an alternative to IEEE 802.11p, the third generation partnership project (3GPP) included support for V2X communications using long-term evolution (LTE) sidelink communications, \textit{a.k.a.}, LTE-V, LTE-V2X, LTE-V2V, or Cellular-V2X (C-V2X). This paper presents a medium access control (MAC) layer performance comparison of these two key enabling technologies by utilizing discrete-time Markov chains (DTMCs).

The first WiFi-based standard specifically designed for vehicular communications was approved under IEEE 802.11p in 2010 \cite{ieee_802_11_p}. IEEE 802.11p was later included in IEEE 802.11-2012 \cite{ieee_802_11_2012}, now superseded by IEEE 802.11-2016 \cite{ieee_802_11_2016}. The European Telecommunications Standards Institute's intelligent transport systems operating in the 5GHz frequency band (ETSI ITS-G5) \cite{CAM_standard, DENM_standard, b16, b17} has been approved as the European version of the IEEE 802.11p standard, to be used in V2X communication applications.
The LTE sidelink was introduced for public safety device-to-device (D2D) communications in Release 12 as Mode 1 and Mode 2. Release
14 introduced Mode 3 and Mode 4, specifically designed for V2X
communications \cite{b2,b4}. Mode 3 enables direct communication between two vehicles, but the selection and management of the radio resources are taken care of by the cellular infrastructure. In Mode 4 \cite{b4,b5}, vehicles autonomously select and manage their radio resources without any support from the cellular infrastructure. In this paper, we focus on ETSI ITS-G5 802.11p and C-V2X Mode 4 for the performance evaluation.

Both 802.11p and C-V2X Mode 4 have many operational similarities. The vehicles that operate using IEEE 802.11p can exchange data through a direct link without the necessity of joining a basic service set (BSS), thus without the need to wait on the association and authentication procedures to complete before the data transfer. Similarly, in C-V2X Mode 4 \cite{b4,b5}, vehicles autonomously select and manage their resources without any cellular infrastructure support, making it particularly superior to C-V2X Mode 3. Both IEEE 802.11p and C-V2X Mode 4 are highly applicable for V2X based safety applications that cannot depend on a third intermediary for communication due to stringent latency constraints. 

Cooperative awareness messages (CAM) and decentralized environmental notification messages (DENM) are two types of broadcast packets used by both IEEE 802.11p and C-V2X Mode 4 in enabling effective communication and ensuring safety. Information related to cooperative awareness, such as position, dynamics, and attributes, is packed in the periodically transmitted CAM packets\cite{CAM_standard}. Thus, CAM packets have a fixed inter-arrival time.  On the other hand, DENM are event-driven messages, triggered by random events such as sudden human-initiated disturbances to vehicle’s pattern of motion ({\em e.g.,}lane changing, signal violation, emergency braking, road-works), and events caused by weather or nature 
\cite{DENM_standard}. The event-triggered DENM packet may get periodically re-transmitted for added reliability \cite{denm_rep}.

In this paper, we focus on the MAC layer operation of the two competing technologies that are significantly different. The multiple access technique in IEEE 802.11p is the well-known carrier sense multiple access with collision avoidance (CSMA/CA). The contention-based protocol requires a vehicle to sense the medium and check if it is idle before transmitting. A mechanism based on random backoff is executed to reduce the probability of collisions. On the other hand, C-V2X Mode 4 utilizes a distributed sensing-based scheduling protocol called semi-persistent scheduling (SPS) \cite{b5}. Vehicles sense the previous transmissions of all neighboring vehicles to estimate free resources, and accordingly pick a free resource for transmission in a manner that avoids packet collisions.

Several works have recently discussed the two technologies from various perspectives, mainly focusing on the physical (PHY) layer, with some providing performance comparisons as well \cite{com1, com2,com3,com4,com6,com7, com8, com5, com9}. A common conclusion of \cite{com1, com2, com3, com4, com6, com7} is that given a target performance threshold, the PHY layer of C-V2X facilitates extended coverage. However, it has been shown that there are specific scenarios and settings where IEEE 802.11p exhibits similar or even better performance \cite{com5, com9}. 
Since our primary focus is on the MAC layer performance, \cite{sps_pseudo_code,b11,b19} can be considered to be the most related to our work. 
To this end, \cite{sps_pseudo_code} is a simulation study that focuses on improving the SPS algorithm. 
The first analytical model for the MAC layer performance of C-V2X Mode 4 is proposed in \cite{b11}. The paper considers a PHY layer model to capture the effect of the distance between a transmitting node and a receiving node, and the SPS algorithm for resource allocation. Then the authors obtain analytical expressions for the MAC layer performance metrics such as the average packet delivery ratio, and for the probability of four different types of transmission errors in C-V2X Mode 4, as a function of the distance between the transmitter and the receiver. 

Our paper, primarily focuses on analytically modeling the MAC layer protocol of C-V2X Mode 4 and IEEE 802.11p by utilizing DTMCs. A DTMC is a well-known method for modeling a MAC layer protocol \cite{b19,b22,markov,DTMC1, DTMC2} as it facilitates the comprehensive modeling of each step in the protocol using different states and state transitions. The steady-state probability of each state can be used to draw insights on the respective step in the protocol. 
A similar DTMC based analytical model for the MAC layer behavior of IEEE 802.11p can be found in \cite{b19}. The proposed model in this paper improves
the model in \cite{b19} along multiple facets. The main novelty in our DTMC for IEEE 802.11p is the higher resolution. 
The improved resolution allows us to study the whole protocol operation at the $aSlotTime$ level, which is the smallest time unit of 13 $\mu$s defined in the standard. The representation also in turn leads to a fair comparison with our model for C-V2X Mode 4 that can be studied at the smallest time unit in its standard, called the subframe (1 ms). 
Although \cite{b19} considers the backoff process of IEEE 802.11p in their modeling, it does not capture the effect of the arbitration inter-frame spacing (AIFS) duration. This is another improvement in the DTMC proposed in the paper. Thus, the model in our paper is significantly different from the one in \cite{b19}.

Another novel aspect of this paper is the implementation of separate DTMC models for CAM (synchronous) and DENM (asynchronous) packet generation, intending to create a more realistic V2X communication environment. With these novel traffic generators, the system can be modeled for more complex and realistic traffic arrival patterns than the simple packet generators found in the literature.  For example, \cite{b11} utilizes a packet generator with a fixed inter-arrival rate, and \cite{b19} utilizes a simple random packet generator. 
The main contributions of our paper can be summarized as follows:
\begin{itemize}
  \item We provide detailed modeling of the MAC layer protocols of C-V2X Mode 4 and ETSI ITS-G5 IEEE 802.11p utilizing DTMCs. 
  The complete Markov model consists of a state machine each for the two competing technologies, two DTMCs to model the generation of CAM and DENM packets, and a queue model to represent the device level packet queue.
  \item We obtain closed-form expressions for the steady-state probabilities of the DTMCs, which are then used to derive expressions for key performance metrics such as the average delay, the collision probability, and the channel utilization of a vehicular network.
  \item We present an application of the models for a highway scenario to provide further insights and comparisons on the derived performance indicators, through numerical evaluations. 
  In particular, we show that C-V2X Mode 4 exhibits a lower collision probability compared to IEEE 802.11p, but IEEE 802.11p maintains a lower average delay compared to C-V2X Mode 4.
\item Design insights on how the MAC-layer performance of both technologies can be improved are presented. These insights can be utilized for future releases and evolution into new radio V2X (NR-V2X) and IEEE 802.11bd \cite{5g_nr_v2x,5g_nr_survey}.
\end{itemize}

The remainder of the paper is organized as follows. The analytical models and the steady-state solutions are presented in Sections \ref{sec:secII} and \ref{sec:secIII}, respectively. Section \ref{sec:secIV} consists of the performance analysis. The numerical results and discussion follow in Section \ref{sec:V}, and Section \ref{sec:VI} concludes the paper.
\vspace{-.90cm}
\section{Analytical Models}\label{sec:secII}
\begin{figure}[t]
    \centering
    \includegraphics[scale=.4]{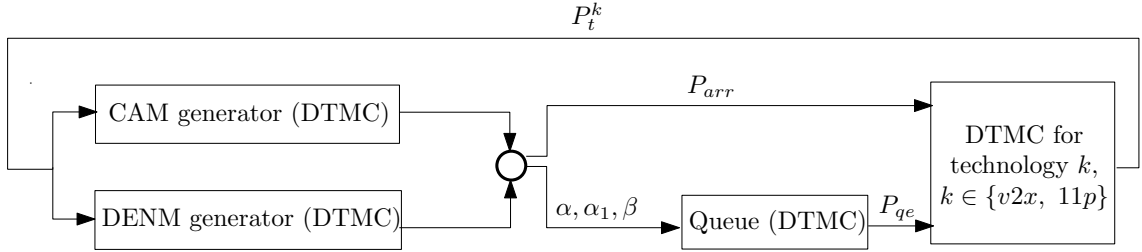}
    \vspace{-10pt}
    \caption{Flowchart illustrating the total model and the dependence among the individual DTMCs.}
    \label{fig:itterative_solve}
    \vspace{-20pt}
\end{figure}

This section presents five DTMCs that are dependent on each other. Firstly, we use two DTMCs to model the generation of CAM and DENM packets. We refer to them as packet generators. The third DTMC models the device level packet queue of a vehicle that consists of the generated CAM and DENM packets. The remaining two DTMCs model the MAC layer operation of C-V2X Mode 4 and ETSI ITS-G5 IEEE 802.11p, respectively.
A holistic view of the overall model that consists of these DTMCs is illustrated in Fig. \ref{fig:itterative_solve}, while also showing how they are interrelated. The parameters that lead to the dependence among the DTMCs will be formally introduced later in the section, while presenting the individual DTMCs. 
All DTMCs ensure that there is a sequence of transitions of non-zero probability from any state to another (irreducible), and that the states are not partitioned into sets such that all state transitions occur cyclically from one set to another (aperiodic). Thus, the DTMCs are ergodic, and hence, a steady-state distribution exists \cite{ergo}. The models are based on non-saturation conditions with regards to transmission, $\ie$, they consider situations when there are no packets to transmit as well, making them more realistic compared to models that assume continuous transmission of packets (saturation conditions). However, the models do not account for a real received power based sensing mechanism. Thus the impact of relative distance, exposed, and hidden terminals are omitted in this study.

\subsection{Packet Generator and Queue Models}

\begin{figure}[t]
    \centering
    \includegraphics[scale=.53]{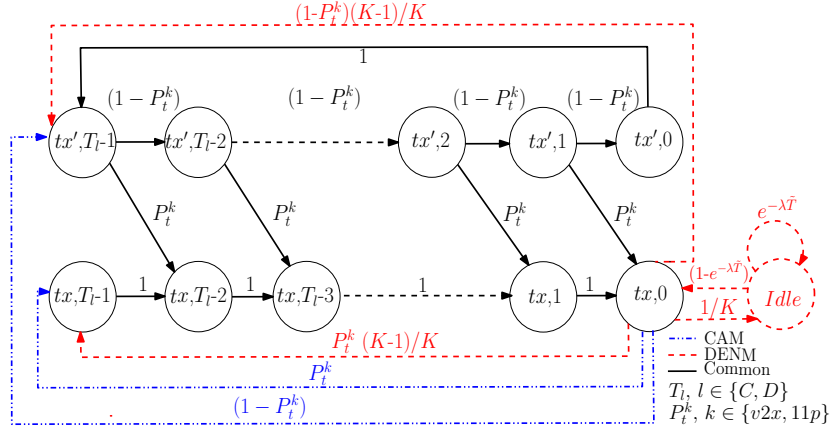}
    \vspace{-10pt}
      \caption{DTMC models for CAM and DENM packet generation, where the line styles are used to differentiate the two models.}
      \vspace{-30pt}
    \label{fig:generator}
    \vspace{-.5cm}
\end{figure}

The generator models of CAM and DENM share significant similarities. Therefore, we use a single figure (Fig. \ref{fig:generator}) to illustrate the DTMCs of the two generator models. The solid lines (black) are used to represent the states and transitions common to the state spaces of both models, and the states and transitions unique to the generation of a particular type of packet are differentiated using line styles and colors. 
The periodic CAM packet generation is modeled using a fixed inter-arrival time model, where the inter-arrival time $T_C$ ms is set in between 100 ms and 1000 ms according to the standard \cite{CAM_standard}. DENM on the other hand are random event-driven messages that are not periodic, thus an additional idle state $(Idle)$ is included in the DENM generator to capture the time periods with no DENM packet generation. A random event triggers the generation of a DENM packet, and by assuming independence among these events, we model the triggered arrivals using a Poisson process of intensity $\lambda$ packet/s. Thus, the probability of at least one DENM packet trigger during $\tilde{T}$ s is given by $1-e^{-\lambda \tilde{T}}$. Due to its critical nature, a DENM packet is repeated $K$ times at a fixed period of $T_D$, for added reliability \cite{denm_rep}. This means, the DENM generator captures two distinct packet types: a Poisson based triggering arrival referred to as {\it trigger} and subsequent fixed-period repetition arrivals referred to as {\it repetition}. The trigger arrival occurs only once per DENM event, and the repetition arrival occurs $K-1$ times following a trigger, periodically, similar to CAM. Due to this reason, the behavior of the CAM arrivals and the DENM repetition arrivals, which are similar in nature, are modeled using the common states. Moreover, according to the standard \cite{DENM_standard}, the originator vehicle has the liberty of setting $T_D$.

The packet generation is represented using states $(i, 0)$, $i \in \{tx,tx^{\prime}\}$. $tx$ and $tx^{\prime}$ is used to differentiate between the current transmit status of the vehicle, $\ie$, whether it has a transmission opportunity, or not, respectively. If it has a transmission opportunity upon generation $(tx, 0)$, the packet is transmitted, followed by a wait of $T_l$ ms, $l \in \{C,D\}$, until the next packet generation. The waiting time is represented with a resolution of 1 ms, referred to as a subframe, by states $(tx, j)$, $j \in \sqparen{0,T_l-1}$. If a transmission opportunity is not available upon generation $(tx^\prime, 0)$, it waits for an opportunity, represented by states $(tx^\prime, j)$, $j \in \sqparen{1,T_l-1}$. A successful transmission results in a state transition from $(tx^\prime, j)$ to $(tx, j-1)$, $j \in \sqparen{1,T_l-1}$. $P_{t}^{v2x}$ and $P_{t}^{11p}$ denote the probability of transmitting a packet in C-V2X Mode 4 and IEEE 802.11p, respectively, and these probability values link the generators with the DTMCs modeling the MAC layer operation, as illustrated in Fig. \ref{fig:itterative_solve}.  

\begin{figure}[t]
\centering
  \includegraphics[scale=.53]{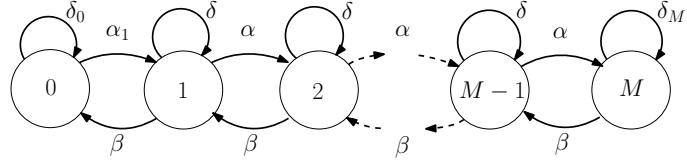}
  \vspace{-10pt}
  \caption{DTMC model for the common packet queue of length $M$ consisting of the generated CAM and DENM packets.}
  \label{fig:queuemodel}
  \vspace{-1.5cm}
\end{figure}

Fig. \ref{fig:queuemodel} illustrates the DTMC that models the device level packet queue per vehicle, consisting of the generated CAM and DENM packets. State $(i)$, $i \in \sqparen{0,M}$, represents a queue length of $i$, where $M$ is the maximum length of the queue. It is not hard to see that the state transitions of this DTMC depend on packet generation and transmission. Thus, the transition probabilities of the queue model are directly related to the packet generators, as shown in Fig \ref{fig:itterative_solve}. 
A transmission of a packet may either lead to maintaining the current state $(i)$ or a state transition from ($i$) to ($i-1$), for $i \in \sqparen{1,M}$, depending on whether a new packet has been generated in the same subframe or not, respectively. Similarly, not being able to transmit a generated packet ($\ie$, traversing through states $(tx^\prime, j)$, $j \in \sqparen{0,T_l-1}$, in Fig. \ref{fig:generator} without a transmission opportunity), leads to a state transition from ($i$) to ($i+1$), for $i \in \sqparen{0,M-1}$. Let $P_{qe}$ denote the probability of the queue being empty, $P_{qne}=1-P_{qe}$, and $P_{arr}$ denote the conditional probability of a new packet arrival given the queue is empty. $P_{qe}$ and $P_{arr}$ links the queue model and the packet generators with the DTMCs modeling the MAC layer operation, respectively, as illustrated in Fig. \ref{fig:itterative_solve}.
\vspace{-.5cm}
\subsection{State Machine for C-V2X Mode 4}
We begin this section by presenting the sensing-based SPS algorithm, which is used for radio resource allocation in C-V2X Mode 4. We follow it up with the respective DTMC model.

\subsubsection{Semi-persistent scheduling algorithm}
The SPS algorithm enables a vehicle to select radio resources without the assistance of an eNodeB, and each vehicle follows the following three steps for resource reservation.

\textit{Step 1: }Within the selection window, which is the time window that initiates with a generation of a packet, vehicle $v$ identifies all possible candidate single-subframe resources (CSRs) that can be reserved. CSRs are groups of adjacent sub-channels within the given 1 ms subframe that are large enough to fit in the sidelink control information (SCI) and the transport block (TB) to be transmitted. The length of the selection window, \textcolor{black}{which is denoted by $\Gamma$}, is defined in the standard as the maximum latency \textcolor{black}{in ms} \cite{b4}, and a CSR should be selected within this duration.

\textit{Step 2: } Based on the information received in the previous 1000 subframes (sensing window), vehicle $v$ creates list $L_1$ that consists of CSRs that it can reserve. $L_1$ includes all the CSRs in the selection window except the ones that satisfy the following conditions.

\begin{enumerate}

  \item CSRs used by vehicle $v$ during the sensing window. This is done as a precautionary measure due to vehicle $v$ not being able to sense these CSRs during its half-duplex transmissions.  
  \item CSRs that are being used by other vehicles at the time vehicle $v$ tries to utilize them (which are known thanks to the information contained in the SCI), and have a  received signal strength indicator (RSSI) value above a threshold level $l_{th}$.     

\end{enumerate}
\noindent
If $L_1$ contains more than 20\% of the total CSRs identified in \textit{Step 1}, the system moves to \textit{Step 3}. Otherwise, $l_{th}$ is increased by 3 dB and \textit{Step 2} is repeated.

\textit{Step 3: }From $L_1$, vehicle $v$ filters out the CSRs that experience the lowest average RSSI values, where the averaging is done over the previous 10 subframes. These CSRs are added to a new list $L_2$ such that the size of $L_2$ amounts to 20\% of the total CSRs in the selection window. Vehicle $v$ randomly and uniformly selects a CSR in $L_2$ and reserves it for the next $RC$ transmissions, where $RC$ denotes the value of the resource counter. Let $RC_F \in \sqparen{R_{l}, R_{h}}$ denote the starting value of the resource counter, where $R_{h}$ and $R_{l}$ are upper and lower limits of $RC_F$, respectively. $RC$ is decremented by 1 for each transmission of a packet, \textcolor{black}{ which happens periodically every $\Gamma$ ms until $RC$ reaches 1}.  When $RC=1$, new CSRs should be selected and reserved with probability $(1-P_{rk})$, where $P_{rk} \in [0, 0.8].$ \textcolor{black}{ This can be done by generating a number randomly and uniformly in $(0,1)$, and then comparing it with the predefined value of $P_{rk}$. Vehicle $v$ continues using the same CSR if the generated random number is less than $ P_{rk}$, and it continues using the subframes encountered in intervals of $\Gamma$ ms for the subsequent transmissions. Else, vehicle $v$ selects a new CSR for the next transmission from $L_2$. Upon new CSR selection, the vehicle randomly and uniformly selects a subframe that falls within the next $\Gamma$ ms for the next transmission. Please refer to \cite{sps_pseudo_code} for a pseudo-code of this algorithm.}

\subsubsection{DTMC model}
\begin{figure}[t]
    \centering
    \includegraphics[scale=.48]{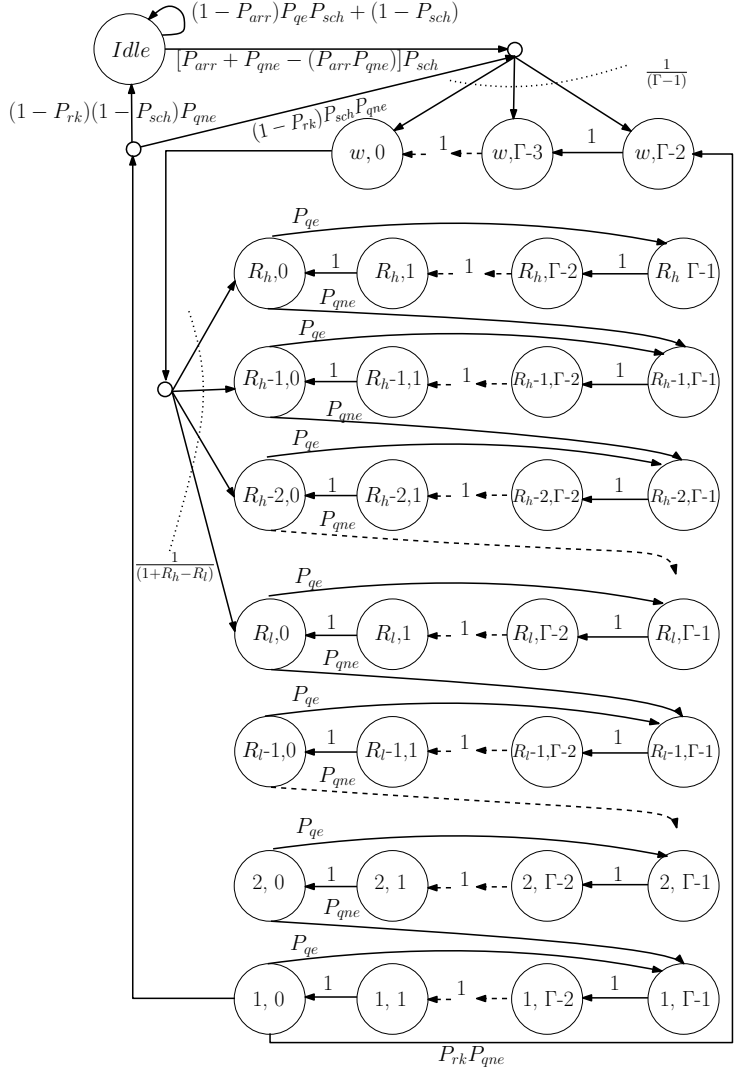}
    \vspace{-10pt}
      \caption{The DTMC modeling the MAC layer behavior of C-V2X Mode 4.}
    \label{fig:state_machine}
    \vspace{-2cm}
\end{figure}

Fig. \ref{fig:state_machine} illustrates the DTMC model for C-V2X Mode 4 operation. 
The state-space of the model is denoted by $S^{v2x}$. Let $P_{sch}$ denote the probability of allocating a suitable CSR for a vehicle through \textit{Steps 1-3}.
State $(Idle)$ represents the state with no packets to transmit, or no CSRs to transmit. 
According to the standard \cite{b2}, there are three selection window sizes with respective ranges for $RC_F$. To this end, the standard includes $\Gamma=100$ ms with $RC_F\in [5, 15]$, $\Gamma=50$ ms with $RC_F\in[10, 30]$ and $\Gamma=20$ ms with $RC_F\in[25, 75]$.

Consider the arrival of a new packet while vehicle $v$ being idle. This necessitates the allocation of a CSR utilizing the SPS algorithm. As $\Gamma$ ms is the maximum allowable latency, the transmission should happen within the next $\Gamma$ subframes. Upon allocation of the CSR, vehicle $v$ selects (randomly and uniformly) a subframe for transmission. Thus, the waiting time before the transmission is modeled by assuming $\Gamma-1$ equiprobable states $(w,\ j)$, where $j\in [0, \Gamma-2]$. As the waiting time elapses, vehicle $v$ selects a value for $RC$ randomly and uniformly from the set of $\paren{1+R_h-R_l}$ values. At every state $(i,0)$, where $i \in \sqparen{1, R_h}$, there is a transmission opportunity, and $i$ represents the current $RC$ value. The device utilizes this opportunity to transmit the control information related to its persistent scheduling. If the queue is not empty, the transmission opportunity is also utilized for data transmission, $i$ is decremented, and the vehicle waits for the next transmission opportunity that arises in $\Gamma$ ms. This waiting time is represented by states $\paren{i-1,j}$, where $j\in[1,\Gamma-1]$. On the other hand, if the queue is empty, the vehicle similarly waits $\Gamma$ ms for the next transmission opportunity. We consider that the vehicle maintains the same $RC$ value $i$ during this waiting period\footnote{As found often in standardization, the standard does not specifically describe what needs to happen to the $RC$ value in such a scenario. Any realization that fulfills the requirements of the standard is deemed to be correct. Note that both decrementing the $ RC $ value or maintaining the same $ RC $ value during this waiting period satisfy the standard requirements. We have used the latter for our model.}. This process repeats until the system reaches state $(1,0)$.

If the queue is still not empty at state $(1,0)$, the vehicle has the option of using the same CSR (with probability $P_{rk}$), or choose a new CSR. If the same CSR is used, the vehicle waits for the maximum waiting time of $\Gamma -1$ ms before choosing a $RC$ value and transmitting. The state transitions for selecting a new radio resource are similar to the transitions described for packet arrival while vehicle $v$ being idle. It is not hard to see that selecting a new radio resource may lead to a lower transmission delay due to the possibility of lower waiting time.

\vspace{-.5cm}
\subsection{State Machine for IEEE 802.11p}
\begin{figure}[t]
\begin{center}
  \includegraphics[scale=.42]{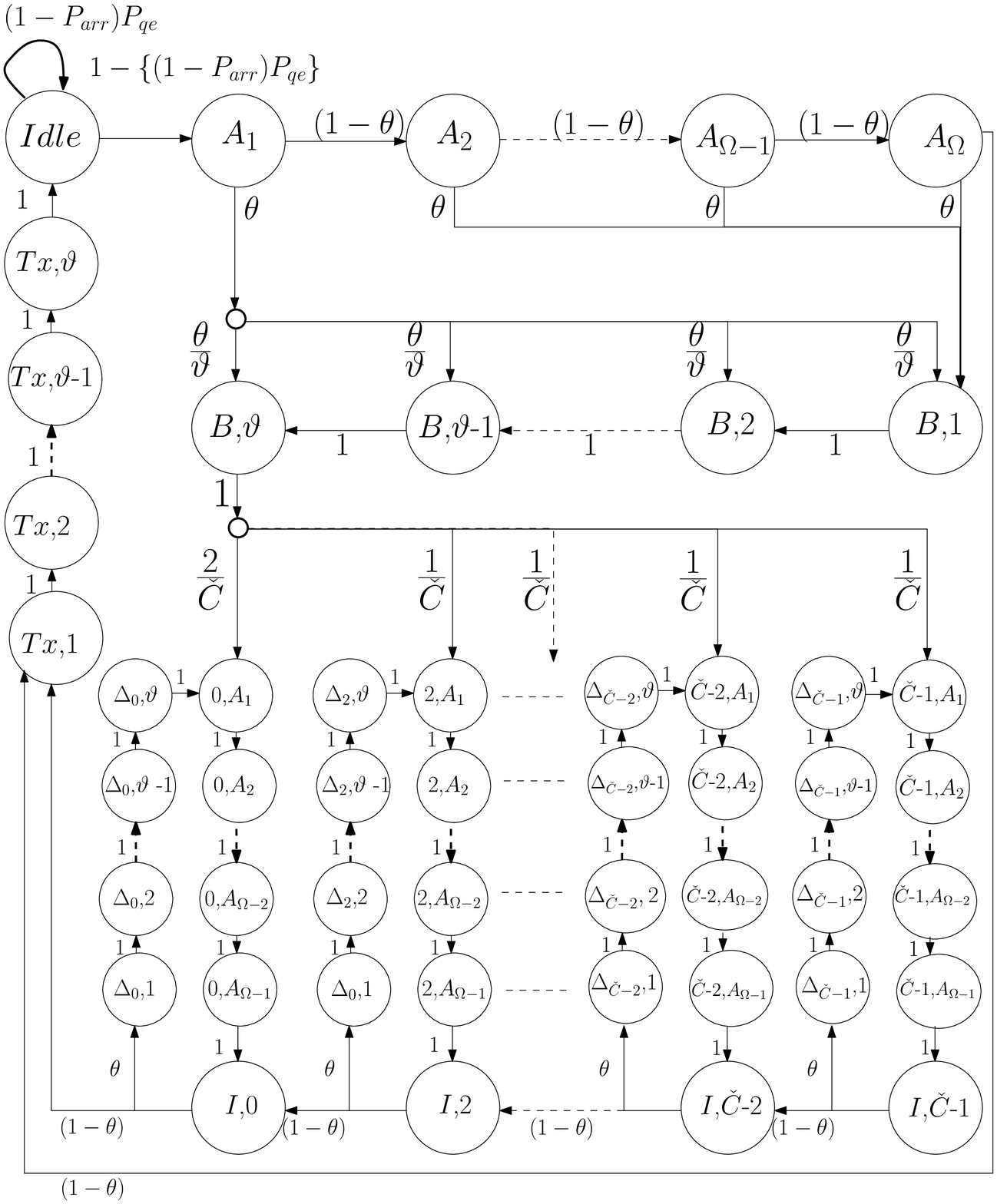}
  \vspace{-10pt}
  \caption{The DTMC modeling the MAC layer behavior of IEEE 802.11p.}
  \label{fig:802_11p_G5_state_machine}
    \end{center}
    \vspace{-1.95cm}
\end{figure}

The state space of the model is denoted by $S^{11p}$ and the state machine is shown in Fig. \ref{fig:802_11p_G5_state_machine}. $\check{C}$ denotes the minimum contention window size. State $(Idle)$ represents the state where there are no packet arrivals, thus the queue is empty. If a packet arrives while being idle, the MAC protocol listens for an $AIFS$ duration before transmitting. The $AIFS$ duration is calculated according to $AIFS=aSIFSTime+AIFSN*aSlotTime$, where $aSlotTime$ is 13 $\mu$s and $aSIFSTime$ is 32 $\mu$s. 
The $AIFSN$ value is selected according to the access category (AC). ETSI specifications do provide four ACs: background, best effort, video, and voice. In this paper, we assume that both CAM and DENM packets utilize the best effort AC. We thus have $AIFSN=6$ and $\check{C}=15$ according to the standard \cite{b16}.

States $(A_i)$ for $ i \in \brparen{1,\dots,\Omega},$ represent the $AIFS$ waiting time, and $\Omega$ denotes the maximum number of $aSlotTime$ intervals per $AIFS$ duration. $\theta$ represents the probability of the channel being busy (channel busy ratio). If the channel is found idle for an $AIFS$ duration, the vehicle is allowed to transmit. Data transmission is represented by states $(Tx,i)$, where $i \in \brparen{1,...,\vartheta }$, and $\vartheta $ denotes the number of $aSlotTime$ intervals required to transmit a packet of 134 bytes over a 6 Mbps control channel (CCH) \cite{b17}. 

If the channel becomes busy during the $AIFS$ duration, the vehicle waits for $\vartheta \times aSlotTime$, which is the time taken for data transmission, until the channel is free again. Waiting is represented by states $(B,i)$, where $i \in \brparen{1,\dots,\vartheta}$. The channel is busy at state $A_1$ depicts a scenario where the packet arrival of the vehicle of interest has occurred while the channel is busy, \textit{i.e.}, another vehicle is transmitting. Thus, the time it has to wait before sensing a free channel is given by $\varrho \times aSlotTime$, where $\varrho$ is a uniformly distributed random integer in $[1,\vartheta]$.  

When the channel becomes free again, that is at state $(B, \vartheta)$, vehicle $v$ initiates a backoff process. The backoff counter value is selected randomly (uniformly) in  $\sqparen{0,\check{C}}$, and the backoff stage is selected depending on the respective backoff counter value. According to the standard \cite{b16}, backoff counter value 0 and 1 both lead to backoff stage 0. Thus, the probability of selecting backoff stage $0$ is twice the probability of selecting any other backoff stage. vehicle $v$ waits for another $AIFS$ duration before sensing the channel again. For backoff counter value $i \in\brparen{0,\dots,\paren{\check{C}-1}}$, states $(i, A_j)$, where $j \in \brparen{1,\dots, \paren{\Omega-1}}$, represent the waiting duration, and $(I, i)$ represent the sensing states. If the channel is found busy at state $(I, i)$, vehicle $v$ waits for $\vartheta \times aSlotTime$, which is represented by states $(\Delta_i,j)$, where $j \in \brparen{1,\dots,\vartheta}$, and another $AIFS$ duration at the same backoff stage $i$. This loop continues until the channel is found idle at state $\paren{I,i}$.  The backoff counter is decremented when the channel is found idle, which takes us to state $(I,i-1)$.  If vehicle $v$ finds the channel to be free at state $(I,0)$, it transmits data.
\vspace{-.25cm}
\section{Steady-state solutions}\label{sec:secIII}
\vspace{-.25cm}
Steady-state solutions of the DTMCs are presented in this section. Firstly, we focus on the CAM and DENM packet generators and the device level packet queue. By utilizing these results, we present the steady-state solutions of the state machines developed for C-V2X Mode 4 and IEEE 802.11p.
\vspace{-.5cm}
\subsection{Queue Model}
We have already discussed the importance of $P_{qe}$, as it links the DTMCs modeling the MAC layer operation with the DTMC modeling the device level packet queue. This probability value can be obtained through the steady-state probability of state $(0)$ in the queue model. From Fig. \ref{fig:queuemodel}, the steady-sate probability of state $(0)$ can be written as
\begin{equation}
 \pi_{0}=\sqparen{1+\alpha_{1}\paren{ \frac{1-\beta^{-M}\alpha^{M}}{\beta -\alpha} }}^{-1}=P_{qe}.
\end{equation}
To obtain  $\alpha$, $\alpha_1$, and $\beta$, we need the steady-state solutions of the CAM and DENM generators, as shown in Fig. \ref{fig:itterative_solve}.

Let $\pi_{i,j}^{C}$ and $\pi_{i,j}^{D}$ denote the steady-state probabilities of state $(i,j)$ of the CAM generator and the DENM generator, respectively. 
To this end, the steady-state probabilities of the CAM generator are given by 
\begin{equation}
\pi_{i,j}^C=
    \begin{dcases}
     \sqparen{ \frac{T_C\sqparen{1-P_{t}^{k}\paren{1-P_{t}^{k}}^{T_C-1}}}{1-\paren{1-P_{t}^{k}}^{T_{C}-1}}}^{-1}&\text{ for }i \in\{tx\},\ j =0\\
    P_{t}^{k}(\pi_{tx,0}^{C}+\sum_{l=j+1}^{T_{C}-1}\pi_{tx^{\prime},l}^{C})&\text{ for }i \in\{tx\},\ j \in [1, T_{C}-2]\\
     \pi_{tx,0}P_{t}^{k}&\text{ for }i \in\{tx\},\ j =T_{C}-1\\
     \pi_{tx,0}^{C}\frac{(1-P_{t}^{k})^{T_{C}-j}}{[1-(1-P_{t}^{k})^{T_{C}-1}]}&\text{ for }i \in\{tx^{\prime}\},\ j \in [0, T_{C}-1]
    \end{dcases}.
    \nonumber
\end{equation}


Similarly, the steady-state probabilities of the DENM generator are given by 
\begin{equation}
    \pi_{i,j}^{D}=
    \begin{dcases}
    \Bigg[\paren{1-\frac{1}{K}}\frac{T_D\sqparen{1-P_{t}^{k}\paren{1-P_{t}^{k}}^{T_D-1}}}{1-\paren{1-P_{t}^{k}}^{T_{D}-1}}+\frac{1}{K}+\frac{1}{K(1-e^{-\lambda \tilde{T}})}\Bigg]^{-1}\\
    \hspace{175pt}\text{ for }i \in\{tx\},\ j =0\\
     P_{t}^{k}\Big[\big(1-\frac{1}{K}\big)\pi_{tx,0}^{D}+\sum_{l=j+1}^{T_{D}-1}\pi_{tx^{\prime},l}^{D}\Big]\hspace{20pt}\text{ for }i \in\{tx\},\ j \in [1, T_{D}-2]\\
     \pi_{tx,0}^{D}\paren{1-\frac{1}{K}}P_{t}^{k}\hspace{85pt}\text{ for }i \in\{tx\},\ j =T_{D}-1\\
     \pi_{tx,0}^{D}(1-\frac{1}{K})\frac{(1-P_{t}^{k})^{T_{D}-j}}{[1-(1-P_{t}^{k})^{T_{D}-1}]}\hspace{15pt}\text{ for }i \in\{tx^{\prime}\},\ j \in [0, T_{D}-1]\\
    \end{dcases}.
    \nonumber
\end{equation}


Since we are using a single queue for both CAM and DENM packets, the transition probabilities of the queue model in Fig. \ref{fig:queuemodel} depend on both generator models that run simultaneously. For $x \in \brparen{\alpha, \alpha_{1}, \beta}$, let $x^C$ and $x^D$ denote the resulting transition probability if only the CAM generator or the DENM generator is in operation, respectively. For two events $E_1$ and $E_2$, $\PR{E_1 \cup E_2}=\PR{E_1}+ \PR{E_2}- \PR{E_1 \cap E_2}$. Thus, $x=x^C+x^D-x^Cx^D$ for $x \in \brparen{\alpha, \alpha_{1}, \beta}$.
To this end, $\alpha^{C}=\pi_{tx^{\prime},0}^{C},\ \alpha_{1}^{C} = \pi^{C}_{tx,0} (1-P_{t}^{k}),\ \beta^{C}=\sum_{j=1}^{T_{C}-1}\pi_{tx^{\prime},j}^{C}P_{t}^{k}$, $\alpha^{D}=\pi_{tx^{\prime},0}^{D},\ \alpha_{1}^{D} = \pi_{tx,0}^{D}\paren{1-\frac{1}{K}} (1-P_{t}^{k}),\  \text{and}\ \beta^{D}=\sum_{j=1}^{T_{D}-1}\pi_{tx^{\prime},j}^{D}P_{t}^{k}$, for $k \in \brparen{v2x,11p}$. 
With similar reasoning, $P_{arr}=\pi_{tx,0}^{C}+1-e^{-\lambda \tilde{T}}-\pi_{tx,0}^{C}\paren{1-e^{-\lambda \tilde{T}}}$.

\subsection{State Machine for C-V2X}
According to the 3GPP C-V2X standard \cite{b4}, single-carrier frequency-division multiple access (SC-FDMA) is considered for the uplink, using a 10 MHz channel. 50 resource blocks (RB) are allocated for this bandwidth per each slot (half subframe), and hence, one subframe contains 100 RBs. A CSR requires at least 4 RBs to transmit a 100 byte payload, using 64 QAM modulation. Therefore, each 1 ms subframe can hold up to 25 CSRs, and hence, the largest selection window of 100 ms can hold up to 2500 CSRs. 20\% of this is 500, and $l_{th}$ can be fine-tuned until we end up with the required number of CSRs. Thus, the standard itself makes it highly unlikely that a randomly selected vehicle ends up without an allocated CSR. Thus, we consider $P_{sch}=1$ in our study without any loss of generality.
Now that we have obtained $P_{qe}$ and $P_{qne}$, the steady-state solutions of the state machine for C-V2X can be used to obtain $P_{t}^{v2x}$ found in the packet generators.

The steady-state equations of the state machine in Fig. \ref{fig:state_machine} are used to derive expressions for its steady-state probabilities, which are presented next. To this end,  
\begin{equation}
 \textit{ State (Idle): } \pi_{Idle}^{v2x} =  b \pi_{w, 0},  \text{ where}\ b=\frac{(1-P_{rk}) \paren{\frac{1}{P_{sch}} - 1}}{P_{arr} + P_{qne} \left(1 - P_{arr} \right)}
\end{equation}
\textit{ States $(w,i)$: }for $ i \in [0,\Gamma-2]$
\begin{align}
 \pi_{w, i} =  \frac{a\pi_{Idle}^{v2x}\paren{\Gamma-1-i}}{\paren{\Gamma-1}}  +  \Big [ \frac{\paren{\Gamma-1-i}}{\paren{\Gamma-1}}\left(1 - P_{rk}\right) P_{\textit{sch}} +  P_{rk} \Big]\pi_{1,0}P_{qne},   
\end{align}
 where  $a =\paren{P_{arr} + P_{qne} - P_{arr} P_{qne}}  P_{sch}$.
\begin{equation}
\textit{ States $(i,j)$: }
\pi_{i,j}=
\begin{dcases}
 \frac{\pi_{w,0}\paren{R_h-i+1}}{P_{qne}^2\paren{1+R_h-R_l}}\hspace{-10pt} &\text{ for } i\in\sqparen{R_l, R_h} ,\  j\in \sqparen{1, \Gamma- 1}\\
\frac{\pi_{w,0}\paren{R_h-i+1}}{P_{\textit{qne}}\paren{1+R_h-R_l}}\hspace{-10pt}
&\text{for } i \in \sqparen{R_l, R_h},\ j=0\\
\frac{\pi_{w,0}}{P_{qne}}, \hspace{-10pt}&\hspace{-1pt} \text{for }\hspace{-1pt} i \in \sqparen{1,R_l-1},\ j\in \sqparen{0, \Gamma-1}
\end{dcases}.
\end{equation}

Since the sum of probabilities is one, we have 
\begin{align}
\begin{split}
 \pi_{w,0} = \Big[1 -\Gamma + b + \paren{\frac{\Gamma-2}{2}} [ab + 2 P_{rk}+(1-P_{rk})P_{sch}]+\frac{\Gamma (R_{h}+R_{l})}{2P_{qne}}\Big]^{-1}.
\end{split}
\end{align}
$P_{t}^{v2x}$ can be obtained by considering product of the probability of transmission opportunity $P_{txo}$ and $P_{qne}$, where   $P_{txo}=\sum_{j=1}^{R_h}\pi_{j,0}$.
 \vspace{-.5cm}
 \subsection{State Machine for ETSI ITS-G5 based IEEE 802.11p}
Next, we present the steady-state solutions for the state machine in Fig. \ref{fig:802_11p_G5_state_machine}. To this end,  
\begin{align}
\textit{State $(A_i)$:} \text{ for } \textit{$i \in \sqparen{1, \Omega}$ } \pi_{A_i}=\pi_{Idle}^{11p}\sqparen{1-P_{qe}\paren{1-P_{arr}}}\paren{1-\theta}^{\paren{i-1}}.
\label{eq:A_i}
\end{align}
  \begin{align}
 \textit{ States $(B,i)$:} \text{ for } \textit{$i \in \sqparen{1, \vartheta}$ }\pi_{B,i}=\pi_{Idle}^{11p}\sqparen{1-P_{qe}\paren{1-P_{arr}}}\sqparen{\frac{\theta}{\vartheta}i-\paren{1-\theta}^\Omega-\theta+1}.
\label{eq:B,i}
\end{align}
 \begin{align}
 \textit{ States $(\Delta_i,j)$:} \text{ for } \textit{$i \in \sqparen{0, \paren{\check{C}-1}}$, $j \in \sqparen{1, \vartheta}$ }
 \pi_{\Delta_i,j}=\pi_{B,\vartheta}\frac{\paren{\check{C}-i}\theta}{\check{C}\paren{1-\theta}}.
\label{eq:Delta_i_j}
\end{align}
\begin{equation}
\textit{ States $(i, A_j)$: } \pi_{i,A_j}=
\begin{dcases}
\frac{\pi_{B,\vartheta}\sqparen{1+\paren{\check{C}-i-1}\theta}}{\check{C}\paren{1-\theta}}\hspace{-9pt}& \text{{ for }} i \in  \sqparen{2, \paren{\check{C}-1}},\ \hspace{-1pt} j \in  \sqparen{1, \paren{\Omega-1}}\\
\frac{\pi_{B,\vartheta}\paren{2-2\theta+\check{C}\theta}}{\check{C}\paren{1-\theta}}\hspace{-9pt}& \text{{ for }} i=0,\ j \in \sqparen{1,  \paren{\Omega-1}}
\end{dcases}.
\label{eq:A_pi_i,A_j}
\end{equation}
\begin{align}
\textit{ States $(I, i)$:} \text{ for } \textit{$i \in \sqparen{0, \paren{\check{C}-1}}$ }
 \pi_{I,i}=\pi_{B,\vartheta}\frac{\paren{\check{C}-i}}{\check{C}\paren{1-\theta}}.
\label{eq:pi_I,i}
\end{align}
\begin{align}
\textit{ States $(Tx, i)$:} \text{ for } \textit{$i \in \sqparen{1, \vartheta}$ }
 \pi_{Tx,i}=\pi_{Idle}^{11p}\sqparen{1-P_{qe}\paren{1-P_{arr}}}.
\label{eq:pi_Tx,i}
\end{align}

By using that the sum of steady-state probabilities is one, and by substituting for $\pi_{A_i}$, $\pi_{B,i}$, $\pi_{\Delta_i,j}$, $\pi_{i, A_j}$, $\pi_{0, A_i}$, $\pi_{I,i}$ and $\pi_{Tx,i}$ from (\ref{eq:A_i}), (\ref{eq:B,i}), (\ref{eq:Delta_i_j}), (\ref{eq:A_pi_i,A_j}), (\ref{eq:pi_I,i}) and (\ref{eq:pi_Tx,i}), respectively, we can show that
\begin{align}
   &\pi_{Idle}=\Big[1+[1-(1-P_{arr})P_{qe}]\Big[[1-(1-\theta)^{\Omega}]\Big[\frac{1}{\theta}+\frac{(\check{C}+1)\theta\vartheta}{2(1-\theta)}+\vartheta\nonumber\\&+(\Omega-1)\frac{[(\check{C}-2)[(\check{C}-3)\theta+2]+(4-4\theta+2\check{C}\theta)]}{2\check{C}(1-\theta)}+\frac{(\check{C}+1)}{2(1-\theta)}\Big]+\frac{\theta}{2}(1-\vartheta)+\vartheta\Big]\Big]^{-1}.
\end{align}
The obtained solution for $\pi_{Idle}$ can then be used to find all other steady-state probabilities, which can then be used to determine $P_{t}^{11p} =\sum_{i=1}^{\vartheta}\pi_{Tx,i}$ and $\theta=1-(1-\sum_{i=1}^{\vartheta}\pi_{Tx,i})^{\paren{N-1}}$. 
 
\section{Performance Analysis}\label{sec:secIV}
This section focuses on deriving expressions for several useful performance parameters that can be used to compare the MAC layer performance of C-V2X Mode 4 and IEEE 802.11p.
\vspace{-.5cm}
\subsection{Probability of Collision $P_{col}$}
\begin{figure}[t]
    \centering
    \includegraphics[scale=.54]{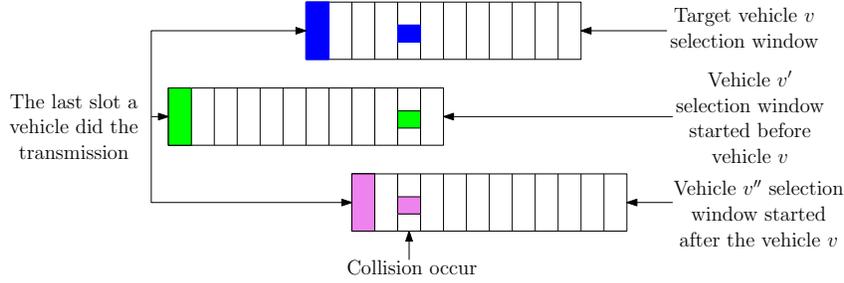}
    \vspace{-10pt}
    \caption{An example for a possible collision in C-V2X Mode 4.}
    \label{fig:vulnarableperiod}
    \vspace{-30pt}
\end{figure}

Even though the SPS algorithm attempts to minimize packet collisions between vehicles at transmission by considering the radio resource utilization of vehicles during the 1000 ms sensing window, there still remains a possibility for collisions. To this end, a schedule collision can occur when a vehicle selects a new radio resource for transmission. In particular, a collision can occur when there is an overlap in the selection windows of neighboring vehicles, as illustrated in Fig. \ref{fig:vulnarableperiod}. In such a scenario, the vehicles with overlap select a CSR independent of each other, and hence, there is a possibility of them selecting the same CSR that leads to collision.  
Let $CSR_{tot}$ and $N$ denote the total number of CSRs in the selection window and the total number of vehicles in the system, respectively. We start the analysis by obtaining an expression for the collision probability of C-V2X Mode 4.
\begin{lemma}\label{lemma:Define_P_col_V2X}
The collision probability of C-V2X Mode 4 is given by
\begin{equation}
P_{col}^{v2x}\approx\ 1\hspace{-2.2pt}-\hspace{-2.2pt}\sqparen{\hspace{-2.2pt} 1\hspace{-2.2pt}-\hspace{-2.2pt}\sqparen{1-\prod_{i=0}^{\Gamma-1}\Big(1-\frac{\pi_{i,0}}{1-\pi_{i,0}i}  \Big)}\frac{(1-P_{rk})}{(CSR_{tot}-N+1)}}^{N-1}.
\label{eq:collisionprob_CV2X}
\end{equation}
\end{lemma}
\begin{IEEEproof}
See Appendix \ref{App:Define_P_col}.
\end{IEEEproof}

The collision probability of IEEE 802.11p is calculated according to $P_{col}^{11p}=1-P_{suc}$, where $P_{suc}$ is the conditional probability that exactly one vehicle transmits on the channel given that at least one vehicle transmits \cite{b22}. An expression for the collision probability of IEEE 802.11p is formally stated through the following lemma. 

\begin{lemma}\label{lemma:Define_P_col_11p}
The collision probability of IEEE 802.11p is given by
\begin{align}
P_{col}^{11p}=1-  {\frac{N\sqparen{\paren{1-\theta}\paren{\pi_{I,0}+\pi_{A_\Omega}}+\sum_{i=1}^{\vartheta}\pi_{T,i}}\sqparen{1-\paren{\pi_{I,0}+\pi_{A_\Omega}+\sum_{i=1}^{\vartheta}\pi_{T,i}}}^{\paren{N-1}}}{1-\sqparen{1-\paren{\pi_{I,0}+\pi_{A_\Omega}+\sum_{i=1}^{\vartheta}\pi_{T,i}}}^{N}}.}
\label{eq:collisionprob_80211p}
\end{align}
\end{lemma}
 \begin{IEEEproof}
See Appendix \ref{App:Define_P_col}.
\end{IEEEproof}
\vspace{-20pt}
 \subsection{Average Delay $D_{ave}$}

Next, we focus on the average delay between the generation and the transmission of a packet. The delay value captures the queuing delay, which is the time a packet waits in the queue making way to the previously generated packets, and the access delay, which is the time a vehicle waits before being able to access the shared radio resources. Firstly, we present an expression for the average delay of C-V2X Mode 4 through the following lemma.
\begin{lemma}\label{lemma:Define_D_V2X}
The average delay of C-V2X Mode 4 is given by
\begin{equation}
d_{avg}^{v2x} = \frac{\sum_{i=1}^{m}\frac{2i-1}{2P_{txo}}\pi_{i}}{1-P_{qe}}. 
\label{Eq:average_delay_C_V2X}
\end{equation}
\end{lemma}
\begin{IEEEproof}
See Appendix \ref{App:Define_Delay}.
\end{IEEEproof}

For IEEE 802.11p, the average delay is calculated by utilizing the delay of each state in the state machine, except the idle state. The normalized average delay of the system is calculated using the delay values of the individual states. Let $D_{i,j}$ denote the delay at state $\paren{i,j}$. $aSlotTime$ is used as the unit delay, thus $D_{T,1}=\vartheta$ since the transmission of a packet of 134 bytes takes $\vartheta \times aSlotTimes$. We assume $D_{I}=0$. To this end, the delay at each state of the system is calculated according to the following equations.
\begin{equation}
\textit{ States $(I, i)$:}  D_{I,i}=
\begin{dcases}
\frac{1+\vartheta+\theta\paren{\Omega-1}}{\paren{1-\theta}} &\text{ for } i=0\\
\frac{i+\vartheta\sqparen{1+\theta\paren{i-1}}+i\theta\paren{\Omega-1}}{\paren{1-\theta}}&\text{ for } i \in \sqparen{2, \check{C}-1}
\end{dcases}.
\nonumber
\end{equation}
\begin{align}
\textit{ States $(i, A_j)$:} \text{ for } \textit{$i \in \brparen{0, 2, \ldots ,\check{C}-1}$, $j \in \sqparen{1, \paren{\Omega-1}}$ }
 D_{i,A_j}=\paren{\Omega-j}+D_{I,i}.\nonumber
\end{align}
 \begin{align}
 \textit{ States $(\Delta_i,j)$:} \text{ for } \textit{$i \in \brparen{0, 2, \ldots ,\check{C}-1}$, $j \in \sqparen{1, \vartheta}$}
 D_{\Delta_i,j}=\paren{\vartheta-j+1}+\paren{\Omega-1}+D_{I,i}.\nonumber
\end{align}
\begin{equation}
\textit{ States $(B,i)$:}
D_{B,i}=
\begin{dcases}
1+\frac{2}{\check{C}}\sqparen{\paren{\Omega-1}+D_{I,0}}+\frac{\paren{\check{C}-2}\paren{\Omega-1}}{\check{C}}+\sum_{i=2}^{\check{C}-1}D_{I,i}\text{ for }i=\vartheta\nonumber\\
1+D_{B,\paren{i+1}}\text{ for }i \in \sqparen{1, \paren{\vartheta-1}}
\end{dcases}.
\end{equation}
\begin{align}
\textit{ States $(Tx, i)$:} \text{ for } \textit{$i \in \sqparen{1, \vartheta}$ } D_{Tx,i}=\vartheta-\paren{i-1}.\nonumber
\end{align}
\begin{equation}
\textit{ States $(A_i)$:}
D_{A_i}=
\begin{dcases}
1+\paren{1-\theta}D_{A_{i+1}}+\frac{\theta}{\vartheta}\sum_{j=1}^{\vartheta}D_{B,j}\hspace{-10pt}&\text{ for } i=1 \nonumber\\
1+\paren{1-\theta}D_{A_{i+1}}+\theta D_{B,1}&\text{ for }i \in \sqparen{2, \paren{\Omega-1}} \nonumber\\
1+\paren{1-\theta}D_{Tx,1}+\theta D_{B,1}\nonumber
 &\text{ for }i=\Omega\nonumber
\end{dcases}.
\end{equation}

By using these equations, the average delay of the system is obtained through the following lemma.
\begin{lemma}\label{lemma:Define_D_11p}
The average delay of IEEE 802.11p is given by
\begin{align}
 d_{avg}^{11p}=\vartheta+1 + \sum_{i=1}^{\Omega-1} (1 - \theta)^i+ \sum_{i \in \brparen{S^{11p} - \brparen{Idle, \bigcup_{j=1}^{\vartheta}Tx,j, \bigcup_{k=1}^{\Omega}A_k}}} \frac{D_i \pi_i}{\sqparen{1 -\paren{\pi_{Idle} + \sum_{j=1}^{\vartheta}\pi_{Tx,j} + \sum_{k=1}^{\Omega}\pi_{A_k}}}}.
\label{Eq:average_delay_802_11p}
\end{align}
\end{lemma}
\begin{IEEEproof}
See Appendix \ref{App:Define_Delay}.
\end{IEEEproof}
\vspace{-.7cm}
\subsection{Average Channel Utilization}
\vspace{-.25cm}
The average channel utilization depicts the average number of users successfully accessing the channel simultaneously. Thus, the average channel utilization of C-V2X Mode 4 and IEEE 802.11p is given by 
\begin{equation}
\vspace{-10pt}
CU_{avg}^{v2x} =\frac{P_{t}^{v2x}N(1-P_{col}^{v2x})}{\textit {CSRs per subframe}}
\label{eq:average_channel_utilization_CV2X}
\end{equation}
and
 \begin{equation}
 \vspace{-10pt}
CU_{avg}^{11p} = P_{t}^{11p}N(1-P_{col}^{11p}), 
\label{eq:average_channel_utilization_80211p}
\end{equation}
respectively. Note that since we are interested in finding the average channel utilization within a single subframe in C-V2X Mode 4, we normalize the channel utilization value by the total number of CSRs within a single subframe, which is given by $CSR_{tot}$/$\Gamma$. 
 
\section{Numerical Results and Discussion}\label{sec:V}
\vspace{-.25cm}
In this section, we present an application of the models for a highway scenario to provide insights and comparisons on key performance indicators through numerical evaluations. We were unable to find or generate similar data from a real vehicular network testbed for validation. 
\vspace{-.5cm}
 \subsection{Instantiation of CAM, DENM and the DTMC Models in a Highway}
 \vspace{-.25cm}
We consider a highway with four parallel lanes in each direction. We assume that the vehicles move at a constant speed of 100 km/h on the highway, and the average inter-vehicle gap is 50 m. \textcolor{black}{Note that since we do not model the locations of the vehicles in Section II, these parameters are only provided to obtain a feasible range for $N$ within the coverage region of vehicle $v$}. We also assume that only CAM and DENM are utilized for V2V communication \cite{b14, b15}, while their reference packet formats are specified according to the ETSI\cite{CAM_standard, DENM_standard}. We consider $T_C$ to be between $100$ ms and $1$ s \cite{CAM_standard}, and $T_{D}$ to be 100, 200 and 300 ms.
In IEEE 802.11p, $T_C$ is regulated under the transmit rate control (TRC) technique of the ETSI ITS-G5 decentralized congestion control (DCC) algorithm, where during periods of high/low utilization, $T_C$ is increased/decreased to manage congestion. This is termed as an adaptive CAM rate in the numerical results. $K$ is set between $1$ and $9$, and we consider $\lambda$ to be 0.2 or 1 packets/s. $M$ is considered to be 10 in the queue model.
The steady-state probabilities of the DTMCs are calculated in parallel, which are then used to calculate the probability values that link the DTMCs as shown in Fig. \ref{fig:itterative_solve}. The probability values are iteratively recomputed until they converge.

\vspace{-.5cm}
\subsection{Performance Comparison of C-V2X Mode 4 and IEEE 802.11p}
\vspace{-.25cm}
In this subsection, we compare the MAC-layer performance of IEEE 802.11p and C-V2X Mode 4 by utilizing the performance parameters.

\subsubsection{Average delay}
\begin{figure}[t]
\begin{subfigure}{.5\textwidth}
  \centering
  \includegraphics[scale=.315]{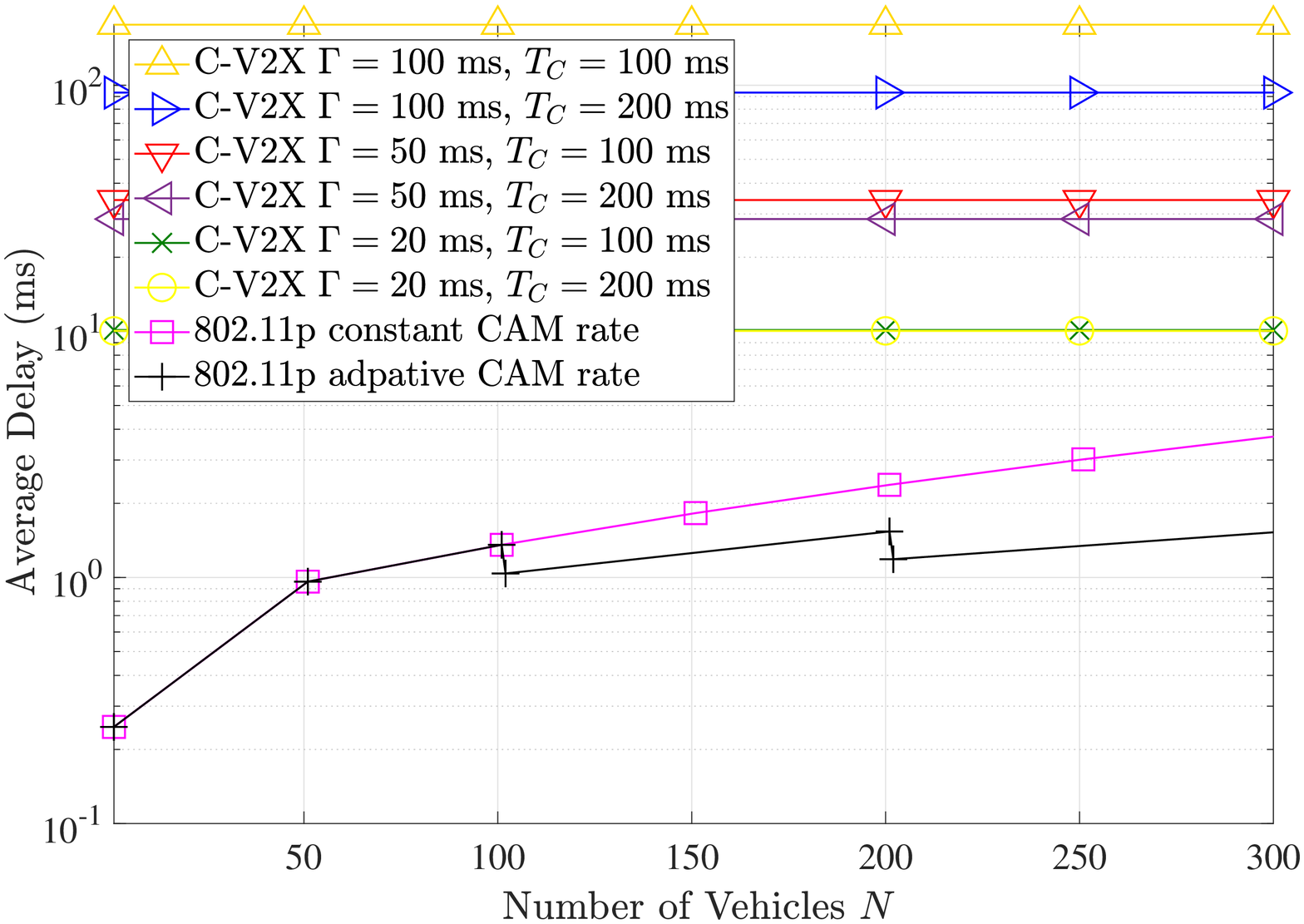}  
  \caption{The average delay vs $N$, where $P_{rk}=0.4$.}
  \label{fig:avg_delay_vs_N_comparison}
\end{subfigure}
\begin{subfigure}{.5\textwidth}
  \centering
  \includegraphics[scale=.315]{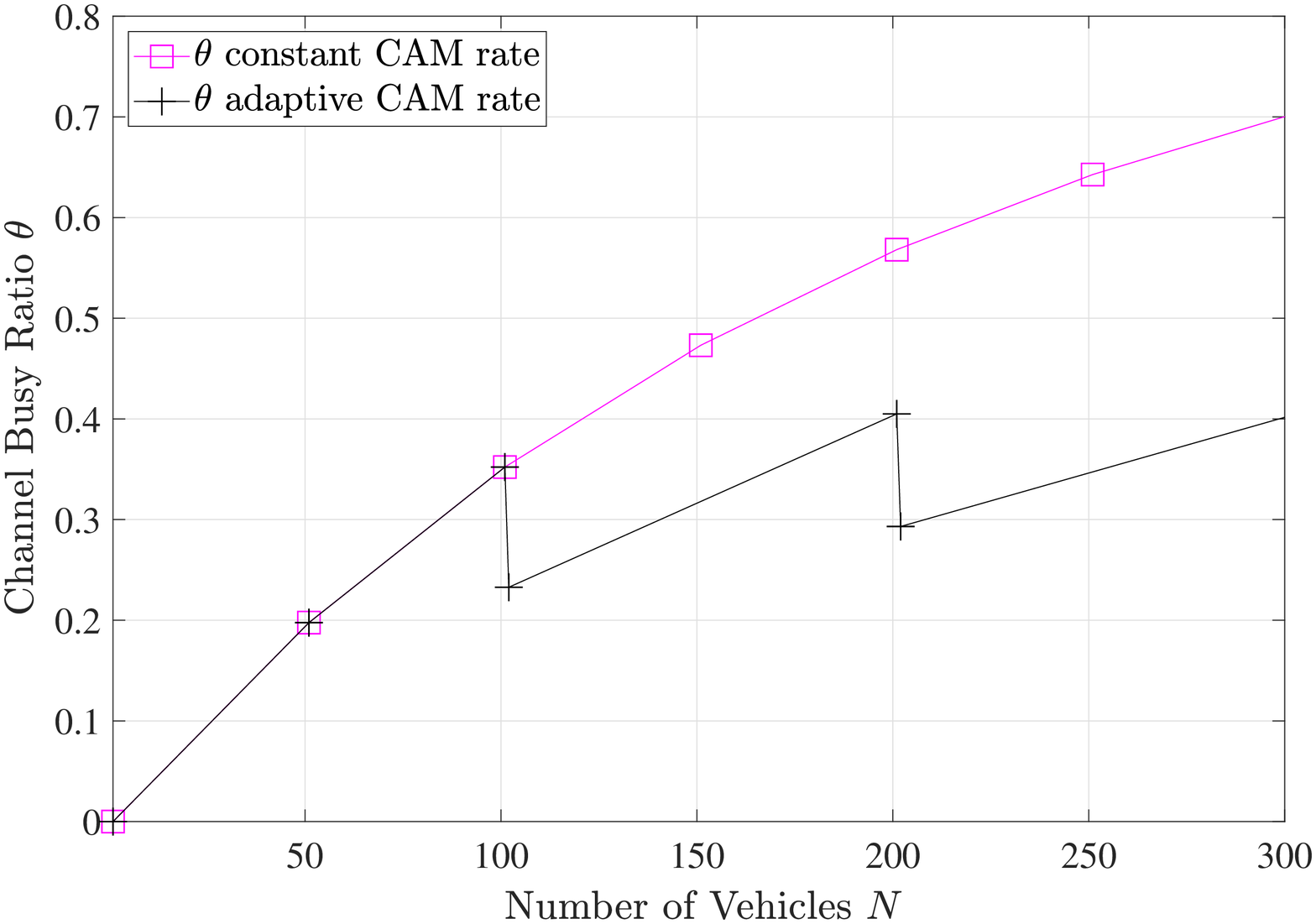} 
  \caption{The channel busy ratio vs $N$, where $T_{C}=100$ ms.}
  \label{fig:tita_vs_N}
\end{subfigure}
\vspace{-10pt}
\caption{The behavior of the average delay and the channel busy ratio with the number of vehicles $N$, where $\lambda= 1$ packet/s, $T_{D}= 100$ ms, and $K=5$.}
\vspace{-1.75cm}
\end{figure}

Average delay is calculated according to $\paren{\ref{Eq:average_delay_C_V2X}}$ and $\paren{\ref{Eq:average_delay_802_11p}}$ for C-V2X Mode 4 and IEEE 802.11p, respectively. Fig. \ref{fig:avg_delay_vs_N_comparison} illustrates the variation of the average delay with $N$.
Firstly, we can observe that IEEE 802.11p is superior to C-V2X Mode 4 in terms of delay. The lower delay in IEEE 802.11p is mainly due to the maximum $AIFS$ duration being 149 $\mu$s \cite{b16}. This is approximately equal to 12 $aSlotTimes$, thus to transmit a 134 byte packet, it takes 14 $aSlotTimes$ over the CCH. Therefore, even after adding the average backoff delay to the above-calculated delay, it is unlikely that the total average delay is greater than few milliseconds. This is much smaller compared 20 ms, which is the smallest selection window size in C-V2X Mode 4, and where it does the best in terms of average delay, as shown in Fig. \ref{fig:avg_delay_vs_N_comparison}. 
We can observe that the average delay increases further with $\Gamma$ in C-V2X Mode 4. For a given value of $\Gamma$, a higher average delay can be observed with a reduction of the CAM inter-arrival time from 200 to 100 ms. This is due to the increased congestion in the network.

It is interesting to note that the average delay is not sensitive to the number of vehicles in C-V2X Mode 4 compared to IEEE 802.11p, where the average delay increases with $N$.  C-V2X uses a scheduling based resource allocation method. We have already discussed that the standard ensures $P_{sch}=1$, thus there exist ample radio resources for all users to transmit. This is the main reason for the flat behavior of the average delay with respect to $N$. On the other hand, IEEE 802.11p resorts to a contention-based access mechanism. Therefore, the delay increases monotonically with $N$. The explanation is consistent with Fig. \ref{fig:tita_vs_N} that depicts the channel-busy ratio $\theta$, which is a metric used to capture the busyness of the channel in IEEE 802.11p, increasing with $N$. The channel is considered to be busy if a vehicle other than the target vehicle is transmitting. It can also be seen from Fig. \ref{fig:tita_vs_N} that the adaptive CAM rate affects the channel busy ratio favorably, and thus helps in reducing the average delay associated with IEEE 802.11p further, as can be observed in Fig. \ref{fig:avg_delay_vs_N_comparison}. 



\begin{figure}[t]
\begin{subfigure}{.5\textwidth}
  \centering
  \includegraphics[scale=.315]{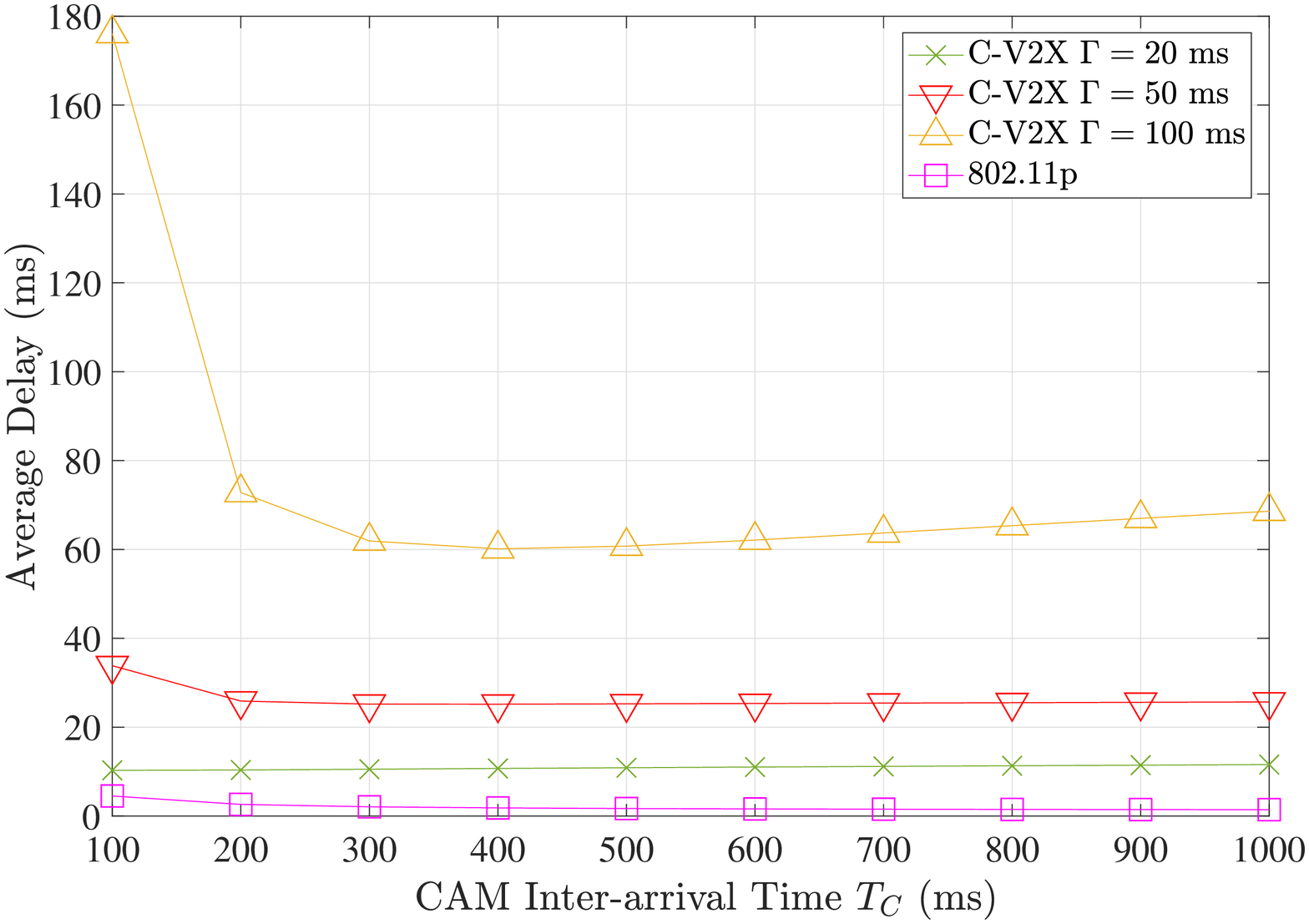}  
  \caption{Average delay vs $T_C$, where $N=300$, $\lambda=1$ packet/s, $T_{D}= 100$ ms, $P_{rk}=0.4$, and $K=5$.}
  \label{fig:delay_vs_T_C}
\end{subfigure}
\begin{subfigure}{.5\textwidth}
  \centering
  \includegraphics[scale=.315]{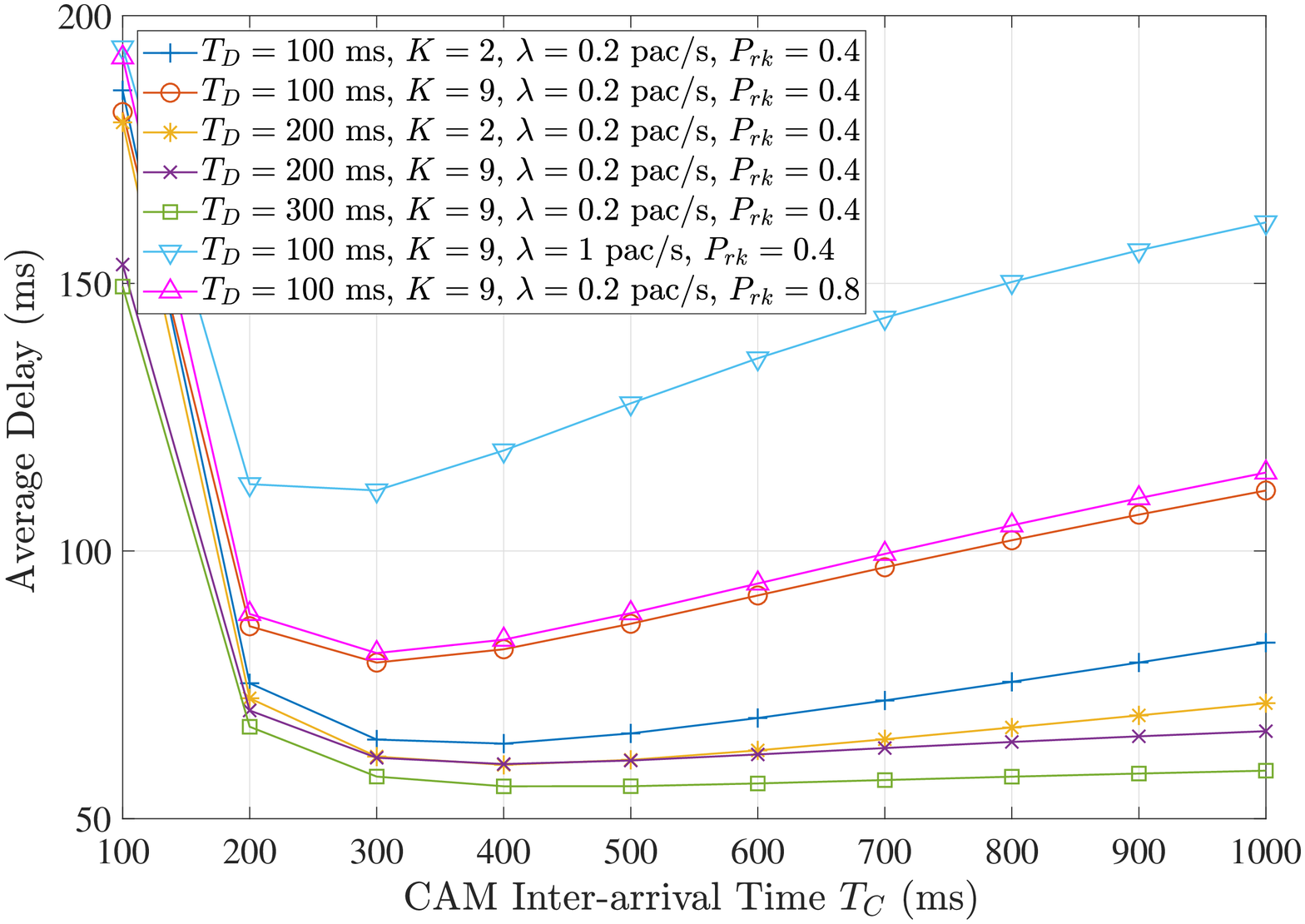}
  \vspace{-30pt}
  \caption{Average delay of C-V2X Mode 4 vs $T_C$ for different combinations of $T_{D}$, $K$, $\lambda$ and $P_{rk}$, where $N=50$, and $\Gamma=100$ ms.}
  \label{fig:delay_vs_t}
\end{subfigure}
\vspace{-10pt}
\caption{The behavior of the average delay with the inter-arrival time of CAM packets $T_C$. }
\vspace{-1.7cm}
\end{figure}

The average delay variation of both technologies with $T_{C}$ is shown in Fig. \ref{fig:delay_vs_T_C}. Intuitively, the behavior with $T_D$ should be similar as the modeling of CAM generation and DENM repetition is identical. Firstly, we can see the average delay reducing with $T_C$ as observed and explained in Fig. \ref{fig:avg_delay_vs_N_comparison}. Secondly, with regards to C-V2X, we can see that $ \Gamma $ has a higher impact on the average delay than $ T_C $. The value of $ \Gamma $ dominates the delay, $\ie$, we cannot negate the adverse effect on the delay caused by an increase in $ \Gamma $ by simultaneously increasing $T_C$. We note that there is a tradeoff in reducing $\Gamma$ as well. While reducing the average delay, it simultaneously reduces the number of CSR values, and hence, it reduces the number of vehicles that can be supported simultaneously. The average delay associated with IEEE 802.11p reduces monotonically with $T_C$. The corresponding behavior associated with C-V2X is more interesting as it first decreases and then increases with regards to $T_C$ (the variation is negligibly small for $\Gamma=20$ ms). Thus, we further elaborate on the variation of the average delay of C-V2X Mode 4 with $T_C$, $P_{rk}$, and different parameter combinations of the DENM generator model such as $T_D$ and $K$ in Fig. \ref{fig:delay_vs_t}.

According to Fig. \ref{fig:delay_vs_t}, we can observe that the average delay increases when $T_D$ decreases, or when the DENM packet arrival rate increases. Both of these observations are due to the congestion caused by more transmissions per unit time. When $T_{D}=100$ ms, the average delay increases with $K$. However, when $T_{D}=200$ ms, the average delay decreases with $K$. This phenomenon can be justified as follows. When $T_{D}=100$ ms, the service rate is nearly equal to the packet repetition frequency. This results in more CAM and DENM packets in the queue, leading to higher queuing delays. However, when $T_{D}=200$ ms, the service rate is higher than the repetition interval of DENM packets. In such a scenario, increasing the average number of repetitions results in the target vehicle encountering the random waiting time, which has a maximum average delay of 50 ms (when $\Gamma=100$ ms), more frequently compared to waiting through the whole resource reservation interval (RRI), which is of 100 ms. This leads to a reduction in average delay. Similar behavior can be observed for $T_{D}=300$ ms as well.



The average delay of a vehicle operating in C-V2X Mode 4 is critical in nature, and need to be minimized for efficient V2X communication. 
An interesting observation in the behavior of the average delay is the existence of a locally optimal point. For an example, when $T_{D}=100$ ms, $K=9$ and $\lambda=0.2$ packets/s, the lowest average delay can be observed at $T_{C}=300$ ms implying the average delay can be reduced further by dynamically changing the CAM packet generation rate based on the packet arrival rate of DENM packets. The DENM packet arrival rate is based on the occurrence of an event and its severity. It can be concluded that the vehicle can reduce the overall average delay further in C-V2X Mode 4 communication if it can change the CAM packet generation rate based on $\lambda$, to achieve the local optimal point of the delay curve shown in Fig. \ref{fig:delay_vs_t}. 

We can also observe the average delay increasing with $P_{rk}$ in Fig. \ref{fig:delay_vs_t}. High values of $P_{rk}$ curtails the vehicle from choosing new radio resources for transmission. When $P_{rk}$ is low, a vehicle again receives more opportunities to encounter the waiting interval (average duration of 50 ms), compared to the longer RRI intervals (duration of 100 ms). Thus, high $P_{rk}$ values lead to higher average delays.

\subsubsection{Collision probability}
\begin{figure}[t]
\begin{subfigure}{.5\textwidth}
  \centering
  \includegraphics[scale=.315]{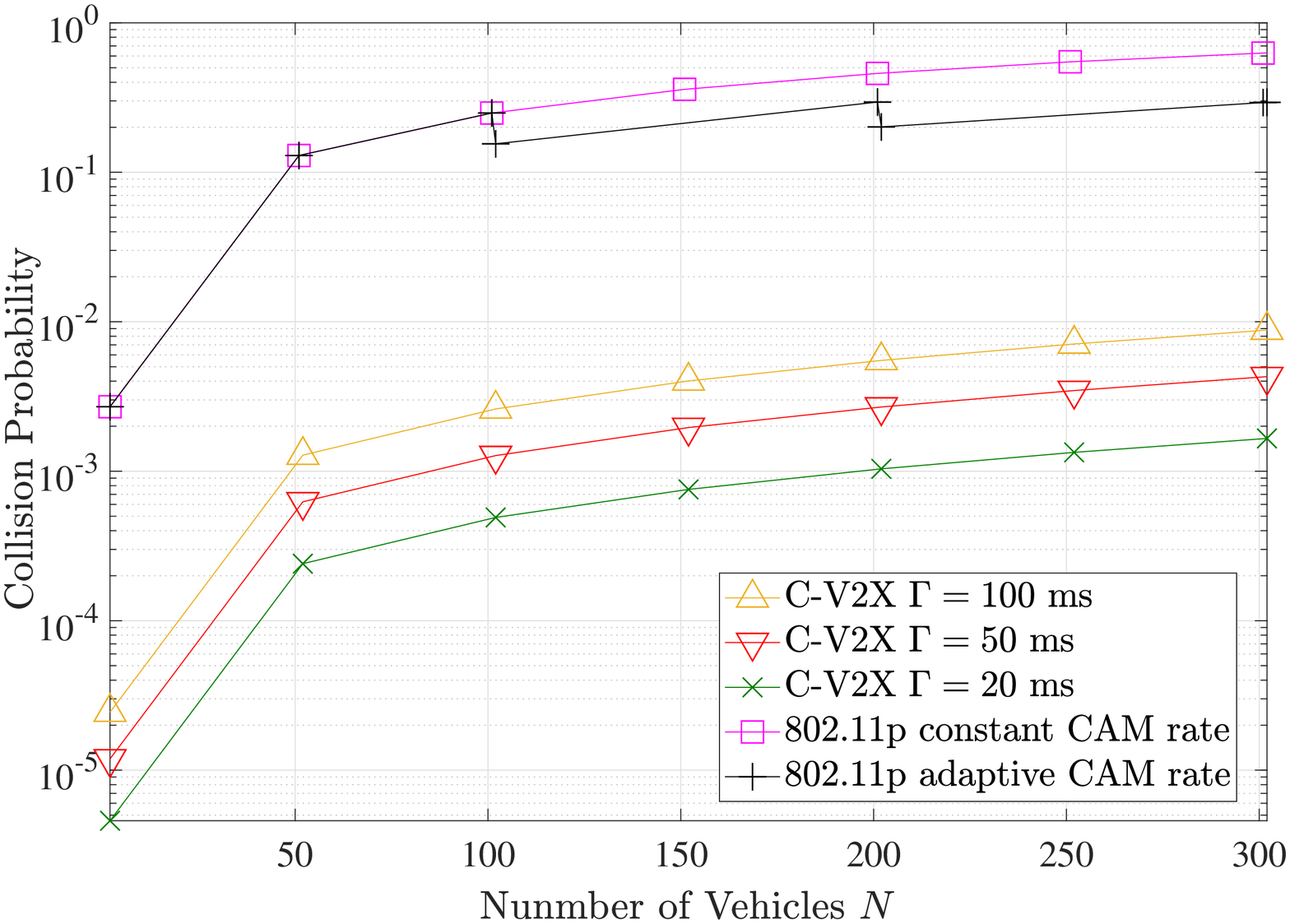}  
  \caption{Collision probability vs $N$.}
  \label{fig:P_col_vs_N_comparison}
\end{subfigure}
\begin{subfigure}{.5\textwidth}
  \centering
  \includegraphics[scale=.315]{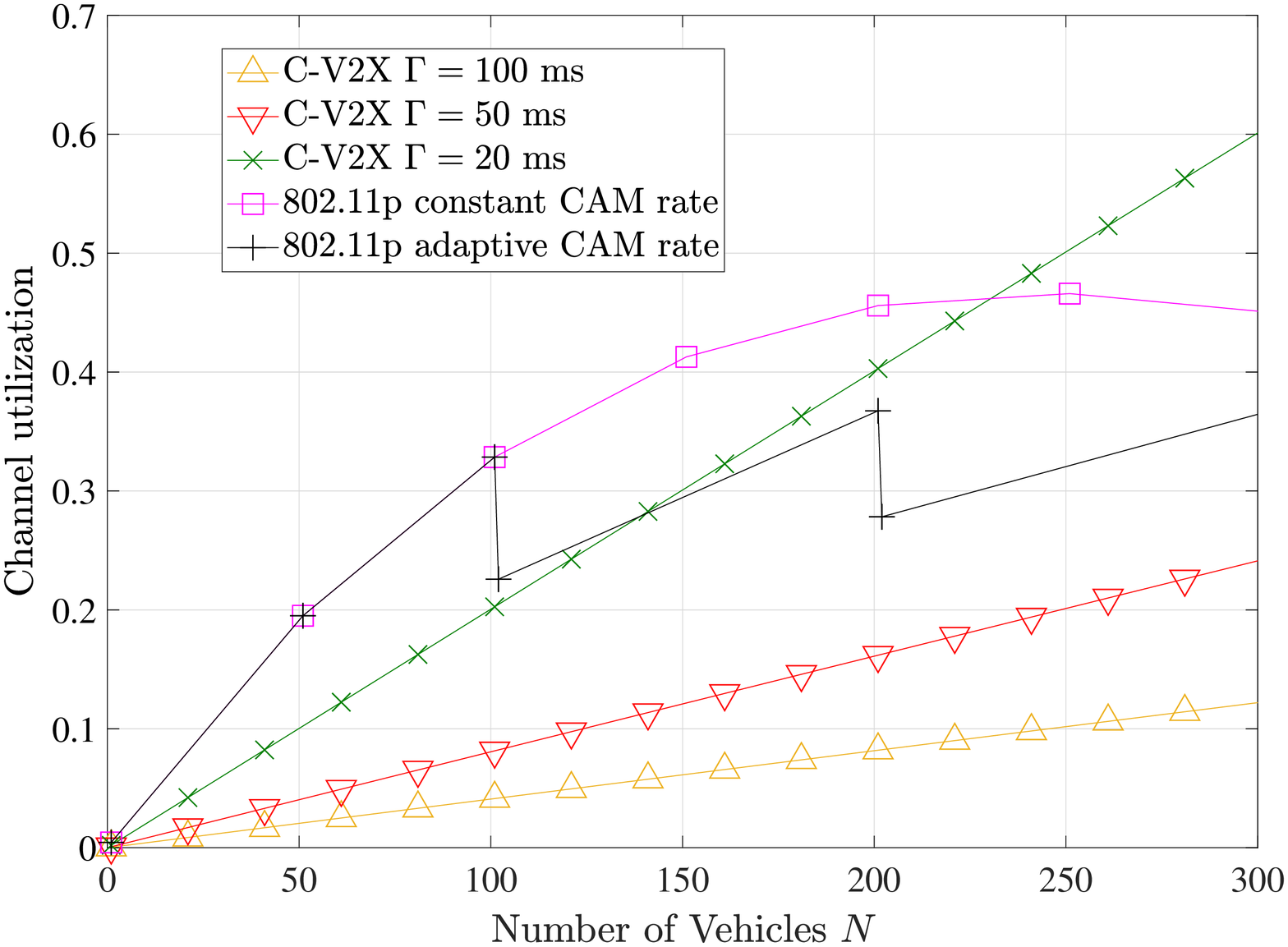} 
  \caption{Average channel utilization vs $N$.}
  \label{fig:channel_utilization _vs_N_comparison}
\end{subfigure}
\vspace{-10pt}
\caption{The behavior of the collision probability and the average channel utilization with the number of vehicles $N$, where $\lambda= 1$ packet/s, $T_{C}= T_{D} =100$ ms, $P_{rk}= 0.4$, and $K=5$.}
\label{fig:fig}
\vspace{-1.75cm}
\end{figure}

As shown in Fig. \ref{fig:P_col_vs_N_comparison}, it is not surprising that the collision probability increases with $ N $ in both C-V2X Mode 4 and IEEE 802.11p. However, a vehicle that utilizes C-V2X Mode 4 has a lower collision probability than a vehicle that utilizes IEEE 802.11p. Thus, it seems that the SPS algorithm performs better in terms of collision resolution compared to the contention-based method in IEEE 802.11p.  We can observe that the collision probability in C-V2X Mode 4 increases with $\Gamma$. Higher values of $\Gamma$ lead to longer selection windows, which increases the chances of two or more selection windows overlapping, as explained with regards to Fig. \ref{fig:vulnarableperiod}. Thus, the collision probability increases with the value of $\Gamma$. It can be observed that the adaptive CAM rate alleviates the collision probability of IEEE 802.11p marginally, but the collision rate is very high when $N > 50$. The behavior of the collision probability with $T_C$ is similar to what was observed for the average delay in Fig. \ref{fig:delay_vs_T_C}. That is, it decreases monotonically with $T_C$ for IEEE 802.11p, and it decreases first and then increases with a local optimum point for C-V2X. However, although the behavior for C-V2X is similar, the variation with $T_C$ in terms of magnitude is negligibly small.   




\vspace{-5pt}
\subsubsection{Average channel utilization}
As illustrated in Fig. \ref{fig:channel_utilization _vs_N_comparison}, the average channel utilization of C-V2X Mode 4 increases almost linearly with $N$. The rate at which the average channel utilization increases, decreases with the value of $\Gamma$. The system also exhibits lower average channel utilization for longer selection window sizes. In IEEE 802.11p, the average channel utilization increases with $N$ up to about 200, and then saturates. 
The average channel utilization of IEEE 802.11p can be improved with adaptive packet arrival, but in general, it is always higher compared to C-V2X Mode 4. The average channel utilization of both technologies decreases monotonically with $ T_C $ as the number of packets transmitted per second reduces when $ T_C $ increases. Similar behavior can be observed in the channel busy ratio. To this end, for $N=300$, $\lambda=1$ packet/s, $T_{D}= 100$ ms, $P_{rk}=0.4$ and $K=5$, the reduction in average channel utilization when $T_C$ is increased from 100 ms to 1 s is 17.95\% for IEEE 802.11p. For C-V2X, this reduction is significantly small, {\em e.g.}, for $\Gamma=100$ ms, the reduction is 2.80\%.      




\vspace{-.35cm}
\section{Conclusions}\label{sec:VI}
\vspace{-.25cm}
This paper has presented multi-dimensional DTMC models to compare the MAC-layer performance of the ETSI ITS-G5 IEEE 802.11p and C-V2X Mode 4, considering CAM and DENM packets proposed for ITS. DTMC based traffic generators and a device level queue model have been used to feed the packets to the aforementioned DTMCs for transmission. Closed-form solutions for the steady-state probabilities of the models have been obtained, and they have been then utilized to derive expressions for key MAC layer-specific performance indicators such as the average delay, the collision probability and the average channel utilization. An application for a highway scenario has been used for numerical results. The results have shown how the performance metrics of each communication technology vary for different parameter selections. When comparing the two technologies, the average delay of C-V2X Mode 4 is comparatively higher than IEEE 802.11p. On the other hand, the collision probability of a vehicle communicating using C-V2X Mode 4 is lower than its counterpart. The results have also shown that the average delay of C-V2X has a locally optimal combination of CAM and DENM packet arrival rates, which can be utilized to reduce delays in C-V2X further. Moreover, the TRC technique of the DCC algorithm can be used to regulate the collision probability and the channel utilization of a vehicle communicating using IEEE 802.11p.

\appendices

\section{Derivations of Collision Probability}\label{App:Define_P_col}

\subsection{Proof of Lemma \ref{lemma:Define_P_col_V2X}}
The selection window initiates at reaching state $(1,0)$, and this is the scenario where a collision can occur. The cycle time of state $(1,0)$ is $1/\pi_{1,0}$. Consider that vehicle $v$ initiated its selection window. The probability of a neighboring vehicle reaching state $(1,0)$ during vehicle $v$'s selection window is given by $1-\prod_{i=0}^{\Gamma-1}\paren{1-\frac{1}{1/\pi_{1,0}-i}}=\tilde{p}$. Similarly, the probability of a neighboring vehicle reaching state $(1,0)$ during vehicle $v$'s selection window and selecting the same CSR as vehicle $v$ is given by $\tilde{p}(1-P_{rk})/(CSR_{tot}-CSR_{exc})$, where $CSR_{exc}$ denotes the number of CSRs excluded according to the SPS algorithm such that there are $CSR_{tot}-CSR_{exc}$ CSRs in $L_1$. Thus, the probability of all $N-1$ neighboring vehicles not selecting the same CSR as vehicle $v$ is given by $\sqparen{1-\tilde{p}(1-P_{rk})/(CSR_{tot}-CSR_{exc})}^{N-1}$, and $1- \sqparen{1-\tilde{p}(1-P_{rk})/CSR_{tot}-CSR_{exc})}^{N-1}$ gives us the collision probability.

 Let $\xi$ denote the ratio between the size of the sensing window and the selection window, and $\Phi$ denote the number of times we encounter $RC=1$ in a given sensing window. Since $\xi\leq 2 R_l$ according to the standard \cite{b5}, we have $\Phi \in \brparen{0,1,2}$. This means, depending on the value of $\Phi$, the vehicle of interest $v$ may either use 1 CSR, 2 CSRs or 3 CSRs. Thus, the average number of CSRs used by vehicle $v$ is given by $P_{1}+2P_{2}+3P_{3}$, where $P_{i}$ is the probability of using $i$ CSR values.   
The number of CSRs used by the neighboring vehicles known through SCI is approximately $N-1$. Hence, $CSR_{exc}\approx N-1+ P_{1}+2P_{2}+3P_{3}$. It is not hard to see that $P_{1}+2P_{2}+3P_{3} \leq 3 << N-1$.
Thus, we consider $CSR_{exc}\approx N-1$, which completes the proof. 
\vspace{-.35cm}
\subsection{Proof of Lemma \ref{lemma:Define_P_col_11p}}
\vspace{-.25cm}
Let $P_{suc}=$ \PR{\text{exactly one vehicle transmits $|$} \text{ at least one vehicle transmits}}, which can be simplified as $P_{suc}=$ \PR{\text{exactly one vehicle transmits}} / \PR{\text{at least one vehicle transmits}}. Successful transmission of a packet by vehicle $v$ can be obtained from the steady-state probabilities of the state machine for IEEE 802.11p as $\sqparen{\paren{1-\theta}\paren{\pi_{I,0}+\pi_{A_\Omega}}+\sum_{i=1}^{\vartheta}\pi_{T,i}}$. Similarly, the probability of the $N-1$ neighbors not transmitting is given by $\sqparen{1\hspace{-2pt}-\hspace{-2pt}\paren{\pi_{I,0}\hspace{-2pt}+\hspace{-2pt}\pi_{A_\Omega}\hspace{-2pt}+\hspace{-2pt}\sum_{i=1}^{\vartheta}\pi_{T,i}}}^{(N-1)}$. Thus, the probability of exactly one vehicle transmitting is given by $N\Big[\paren{1-\theta}\paren{\pi_{I,0}+\pi_{A_\Omega}}+\sum_{i=1}^{\vartheta}\pi_{T,i}\Big]\Big[1-(\pi_{I,0}+\pi_{A_\Omega}+\sum_{i=1}^{\vartheta}\pi_{T,i})\Big]^{\paren{N-1}},$
and the probability of at least one vehicle transmitting is given by $1-\sqparen{1-\paren{\pi_{I,0}+\pi_{A_\Omega}+\sum_{i=1}^{\vartheta}\pi_{T,i}}}^{N}$. The ratio of these probabilities gives us $P_{suc}$, and $P_{col}^{11p}= 1-P_{suc}$ completes the proof.

\vspace{-.35cm}
\section{Derivations of Average Delay}\label{App:Define_Delay}
\vspace{-.5cm}
\subsection{Proof of Lemma \ref{lemma:Define_D_V2X}}
From the steady-state probabilities of the queue model, $1/P_{txo}$ is the duration in milliseconds (cycle time) to serve one packet. For the first packet, we may not spend the total cycle time to serve the packet, as it depends on the state vehicle $v$ is in. Thus, we consider the service time to be $\frac{1}{2P_{txo}}$ (half the cycle time) for the first packet. From the second packet onwards, we add $\frac{1}{P_{txo}}$ to the service time of the previous packet to obtain the delay. For example, the service times of the second and the third packets are calculated as $\frac{3}{2P_{txo}}$ and $\frac{5}{2P_{txo}}$, respectively. We consider a queue of length $M$, and the averaging is done by utilizing the steady-state probability of each state, conditioned on the fact that the queue is not empty.  Thus, the average delay is given by 
$d_{ave}^{v2x} = \sum_{i=1}^{M}\frac{2i-1}{2P_{txo}}\pi_{i}/(1-P_{qe}),$
which completes the proof.
\vspace{-.5cm}
\subsection{Proof of Lemma \ref{lemma:Define_D_11p}}
Since unit time is considered to be $aSlotTime$, the delay associated with the transmit states is $\vartheta$. The delay associated with states $A_i$ where $i \in \brparen{1,\dots,\Omega}$ is $1 + \sum_{i=1}^{\Omega-1} (1 - \theta)^i$. The delay associated with the remaining states, \textit{i.e.},  state $i \in S^{11p} - \brparen{Idle, \bigcup_{j=1}^{\vartheta}Tx,j, \bigcup_{i=1}^{\Omega}A_i}$ can be calculated by utilizing the product of the corresponding delay of each state $(D_i)$ with the steady-state probability of each state conditioned on the fact that $i \in S^{11p} - \brparen{Idle, \bigcup_{j=1}^{\vartheta}Tx,j, \bigcup_{i=1}^{\Omega}A_i}$. Sum of the three delay values completes the proof.

\ifCLASSOPTIONcaptionsoff
  \newpage
\fi

\bibliography{Geeth-bibfile}

\end{document}